\documentclass[RNAAS]{aastex62}
\usepackage{graphicx}
\usepackage{epsfig}
\usepackage{multirow}
\usepackage{footnote}

\begin{document}



\title{\large \bf The first polluted white dwarf from {\it Gaia} DR2: the cool DAZ GaiaJ1738$-$0826} 

\author{\large Carl Melis}
\affil{Center for Astrophysics and Space Sciences, UCSD, CA 92093-0424, USA 
\\ email: cmelis@ucsd.edu}

\author{\large B.\ Zuckerman}
\affil{Department of Physics and Astronomy, UCLA, CA 90095-1562, USA} 

\author{\large P.\ Dufour}
\affil{D\'{e}partement de Physique, Universit\'{e} de Montr\'{e}al, Montr\'{e}al, QC H3C 3J7, Canada}

\author{\large I.\ Song}
\affil{Department of Physics and Astronomy, University of Georgia, Athens, GA 30602-2451, USA}

\author{\large B.\ Klein}
\affil{Department of Physics and Astronomy, UCLA, CA 90095-1562, USA}



\begin{figure}[h!]
 \begin{center}
 \includegraphics[trim={0.1cm 12cm 1.8cm 1.8cm},clip,width=180mm]{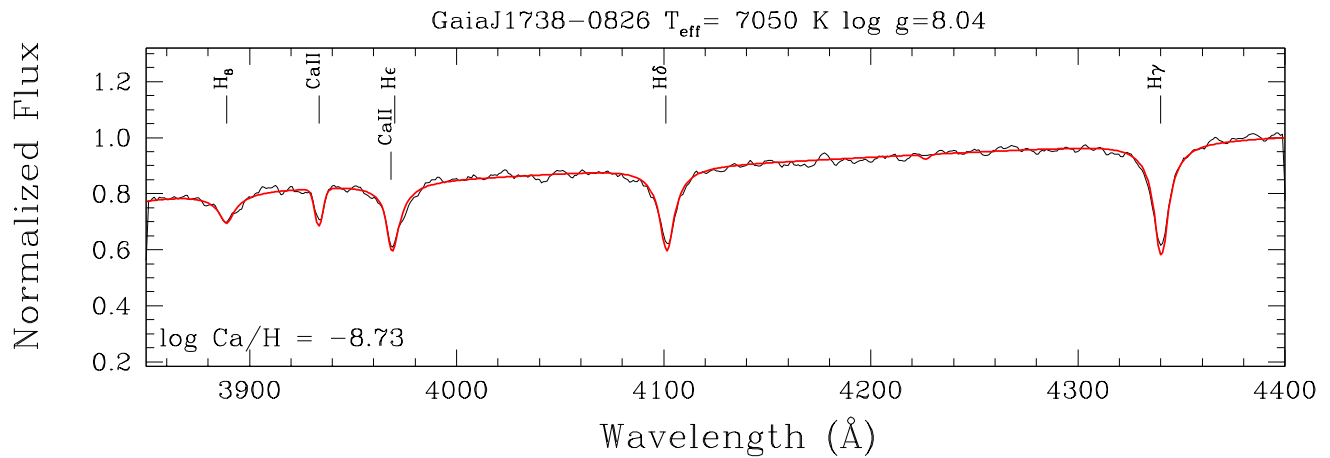}
 \caption{\label{figfits} KAST spectra of GaiaJ1738$-$0826 (black curve) with model having the
 specified parameters overlaid (red line). 
 }
\end{center}
\end{figure}

\large{
\section{}

We present the first metal-polluted single white dwarf (WD) star identified through {\it Gaia} 
DR2 \citep{brown18}. GaiaJ1738$-$0826
was discovered to have strong Ca~II absorption in
initial spectroscopic characterization at Lick Observatory. Notably, GaiaJ1738$-$0826
resembles in many ways the first confirmed metal-polluted hydrogen atmosphere
WD, the DAZ G\,74$-$7 \citep{lacombe83}. 

GaiaJ1738$-$0826 ({\it Gaia} DR2 source 4168312459956062208, 
J2000 RA and DEC 17 38 34.3924 $-$08 26 14.017) was culled from a set of color
and absolute magnitude cuts applied to {\it Gaia} DR2 parallax, G-magnitude, and BP$-$RP
colors. 
Specifically, in a {\it Gaia} color-absolute G-magnitude diagram, WDs were selected to have 
G$_{\rm abs}$ $\geq$ (12.5 $\times$ (BP$_{\rm mag}$ $-$ RP$_{\rm mag}$) + 40) / 4.5.

Characterization spectroscopy was conducted at Lick Observatory with the 3\,m Shane
telescope on UT 22 May 2018. The KAST spectrograph 
(Lick Observatory Technical Reports No.\ 66) was used with light illuminating
a 1$'$$'$ slit and then passed onto the d57 dichroic
which splits light around 5700\,\AA . From there, light is fed into the blue arm of the spectrograph
where it was dispersed with the 600/4310 grism resulting in 1.02\,\AA \,pix$^{-1}$ on the detector.
Light fed into the red arm is dispersed with the 830/8460 grating resulting in 0.94\,\AA \,pix$^{-1}$
on the detector. Spectral coverage is 3450-7800\,\AA .
3000\,seconds total integration time were obtained in the blue and red with final signal-to-noise
ratios of $\approx$35 near 3950\,\AA\ and $\approx$65 near 6500\,\AA . Final spectral resolution
is $\approx$3\,\AA\ in the red and blue. 
Data reduction is done with standard {\sf IRAF} tasks with Feige~34 serving as the standard. 

Spectral modeling is conducted similar to that described in \citet{melis11}.
We began with an initial solution for GaiaJ1738$-$0826 based on fits
to Pan-STARRS PS1 photometry \citep{chambers16} and the {\it Gaia} DR2 parallax
assuming pure H (DA) or pure He (DB) atmospheric compositions;
the spectra show this WD is H-dominated. 
Thus, the H solution was selected 
yielding an effective temperature of 7,050\,K and gravity log$g$ of 8.04 (cgs units).
GaiaJ1738$-$0826 has 
mass 0.6\,M$_{\odot}$, 
radius 0.012\,R$_{\odot}$, 
luminosity 3.3$\times$10$^{-4}$\,L$_{\odot}$, 
cooling time of 1.7\,Gyr, 
and photospheric layer mass ratio
log$q$ of $-$8.3 (e.g., \citealt{dufour17}). A grid of spectra having a range of
Ca abundances is then fit to the Ca\,II H+K lines (see description in \citealt{melis11}).
Figure \ref{figfits} shows a good fit to the Ca and H lines. 
Based on the measured Ca abundance of log$_{10}$[Ca/H]$=$$-$8.7$\pm$0.2,
measured stellar parameters above, and inferred atmospheric diffusion timescale for Ca
of 10$^{3.74}$~years we estimate a Ca mass accretion
rate in steady-state 
of $\approx$2.6$\times$10$^6$\,g\,s$^{-1}$. 
We do not consider thermohaline mixing or changes to convective overshooting 
which, in DA WDs, might dramatically increase derived accretion rates 
(e.g., \citealt{bauer18} and references therein).
Assuming Ca makes up 1.6\% of the 
total heavy element accretion rate (see \citealt{farihi16}), 
GaiaJ1738$-$0826 
could be accreting 1.6$\times$10$^8$\,g\,s$^{-1}$. This value lies below
3$\times$10$^8$\,g\,s$^{-1}$ where WDs frequently have detectable 
infrared excess emission from a circumstellar accretion disk \citep{jura07}.
Archival infrared data from the VISTA Hemisphere Survey and warm {\it Spitzer}
do not reveal the presence of a disk.

In many ways GaiaJ1738$-$0826 resembles the first confirmed DAZ WD, 
G\,74$-$7. G\,74$-$7 was identified from a list of candidate DA,F stars that showed possible
evidence for Ca\,II H+K absorption in early spectroscopic surveys \citep{lacombe83}.
Deeper observations of G\,74$-$7 have detected also Mg, Fe, and Al \citep{zuckerman03}.
Taking the most up-to-date stellar parameters (7,201\,K effective temperature, gravity log$g$
7.93, mass 0.53\,M$_{\odot}$, radius 0.013\,R$_{\odot}$, luminosity
4.1$\times$10$^{-4}$\,L$_{\odot}$, cooling time of 1.4\,Gyr, and photospheric layer mass ratio
log$q$ of $-$8.2), abundances (log$_{10}$[Ca/H]$=$$-$8.83$\pm$0.2,
log$_{10}$[Mg/H]$=$$-$7.90$\pm$0.2,
log$_{10}$[Fe/H]$=$$-$7.89$\pm$0.2,
log$_{10}$[Al/H]$=$$-$9.00$\pm$0.2),
and diffusion timescales
(10$^{3.89}$, 10$^{3.88}$, 10$^{3.88}$, 10$^{3.72}$~years for Mg, Al, Ca, and Fe respectively),
we estimate accretion rates of
7.9$\times$10$^6$, 7.3$\times$10$^5$, 1.5$\times$10$^6$, and 2.7$\times$10$^7$\,g\,s$^{-1}$
for Mg, Al, Ca, and Fe respectively.
Calculating the total heavy element accretion rate as was done for GaiaJ1738$-$0826 above,
we obtain $\approx$10$^8$\,g\,s$^{-1}$.

The first polluted WD, the DZ (helium-dominated atmosphere) 
van~Maanen's Star, was identified in 1917 \citep{vanmaanen17}, 
although it was many years before the significance of this discovery was recognized 
\citep{zuckerman15}.
Polluted white dwarf stars provide a glimpse into the fate of planetary systems and
an unparalleled method of determining the composition of solid material in planetary systems
\citep{zuckerman07};
more recently they have been found to be capable of providing structural information for
massive, differentiated rocky bodies \citep{melis17}. Much in the way the discovery and
confirmation of G\,74$-$7 advanced the study of metal-line WDs by inclusion 
of hydrogen-dominated atmosphere objects $-$ a remarkably delayed 66~years after
van~Maanen's discovery $-$ 
the identification of GaiaJ1738$-$0826 heralds the waiting discovery space amongst
the many thousands of new WDs that {\it Gaia} has made available. 

}

\acknowledgments

Research at Lick Observatory is partially supported by a generous gift from Google.

\end{document}